\documentclass[a4paper]{spie}  

\usepackage{amsmath,amsfonts,amssymb}

\usepackage{graphicx}
\usepackage[colorlinks=true, allcolors=blue]{hyperref}
\usepackage{textgreek} 
\usepackage{support-caption}
\usepackage{subcaption}
\usepackage{rotating}
\usepackage{multirow}
\usepackage{aasmacros}
\usepackage{booktabs}
\usepackage{siunitx}
\usepackage{placeins}

\title{Cryogenic Optical Characteristics of S-FPL51 and S-LAH71: Refractive index, Transmission and Thermal expansion}
\author[a]{J.A.~van~den~Born}
\author[a]{E.~Elswijk}
\author[a]{E.~Dijkstra}
\author[b]{P.~van~Nijnatten}
\author[a]{A.W.~Janssen}
\author[a]{R.~Navarro}

\affil[a]{\footnotesize NOVA Optical Infrared Instrumentation Group at ASTRON, Oude Hoogeveensedijk 4, NL-7991 PD Dwingeloo, The Netherlands}
\affil[b]{\footnotesize OMT Solutions BV, High Tech Campus 9, NL-5656 AE Eindhoven, The Netherlands}

\authorinfo{E-mail: born@astron.nl}

\begin{document} 
\maketitle

\begin{abstract}
Several near-infrared instruments on upcoming extremely large telescopes use a combination of OHARA S-FPL51 and S-LAH71 glass prisms to correct for the atmospheric differential refraction. Because large parts of these instruments are contained within a vessel cooled to liquid nitrogen temperatures, the changes in optical and mechanical properties should be considered during the design phase.

In this report, we provide an update to our continuing efforts to characterize the S-FPL51 and S-LAH71 glass under cryogenic conditions. These efforts include a measurement of the refractive index (293 K, 100 K and 80 K), transmission measurements (293 K, 77 K and 43~K) and a determination of the coefficient of thermal expansion (between 293 K and 45 K). We will describe the cryogenic test setups that we have used to obtain our results, and provide summarizing equations of these material properties.

This characterization will prove useful for conceptual design trade-offs, mechanical interface solutions and detailed optical design for future near-infrared instrumentation on large telescopes.
\end{abstract}

\keywords{CTE, ELT, material properties, MICADO, near-infrared, refractive index, transmission}


\section{INTRODUCTION}
\label{sec:introduction}

Most operational and upcoming infrared instruments on astronomical telescopes feature cryogenically cooled optics, mechanisms and detectors to minimize unwanted thermal radiation and thus increase the signal-to-noise ratio of any observed source. One of the many challenges in the design of these instruments are temperature dependent material properties of the used materials. Although common materials, such as aluminum or fused silica are well characterized for a wide range of environmental conditions\cite{NIST1977_LNG, NIST1994_Fused_Silica, Leviton2006_Fused_silica}, new or highly specialized materials often suffer from a lack of readily available data. If no obvious alternative materials exist, large uncertainties on tolerance calculations or dedicated measurement campaigns are often required as safety margins. In contrast, when accurate data is available, better design trade-offs can be done, detailed mechanical and optical designs can be developed and better predictions on instrumental performance can be performed.  

For near-infrared instruments on the extremely large telescopes, OHARA's low refractive index fluorophosphate glass S-FPL51\cite{OHARA-SFPL51} and high refractive index lanthanum glass S-LAH71\cite{OHARA-SLAH71} have proven to be a popular combination of glasses for the correction of atmospheric differential refraction. MICADO \cite{Davies2022} for the Extremely Large Telescope (ELT) and IRIS \cite{Larkin2016} for the Thirty Meter Telescope (TMT) have both chosen this combination through an extensive search of available glass combinations in order to match the dispersive properties of the atmosphere\cite{vandenBorn2023thesis,Phillips2016}, weighted by instrumental requirements and scientific interests. 

For MICADO, knowledge of the cryogenic refractive index was required for a final iteration of the prism geometry of the Atmospheric Dispersion Corrector (ADC) as well as a more high level optimization of the cold optics design. Accurate information on the thermal expansion differences between the glass and the mount material was needed to ensure that no excessive birefringence would be induced due to high stress in the prism mounting solution. Also, the transmission of the two glasses at their operational temperature had to be determined outside of the range provided by the manufacturer to better define the coating transmission requirements outside of the primary wavelength range of the instrument (0.8 - 2.4~\textmu m), but where the detector will still be sensitive. In addition, there are some materials such as germanium, that show changing transmission profiles at increasing temperatures\cite{Stolberg2015, Thornton1989}. And although we were not aware of materials that vary in transmission at cryogenic conditions compared to room temperature, it would not be detrimental to verify this assumption for S-FPL51 and S-LAH71 substrates.

In this proceeding, we add to the existing and ever growing library of material properties by describing our measurements of the S-FPL51 and S-LAH71 materials. Because of the above mentioned reasons, we have carried out different measurement campaigns to determine the refractive index, transmission profile and coefficient of thermal expansion at both room temperature and at cold cryogenic conditions. The measurements were done on samples that were produced from the same manufacturing run as the glass that will be used for the final hardware in MICADO. For these two materials we will present the temperature dependent refractive indices in section~\ref{sec:ref_index}, the (cryogenic) transmission profile between 500 and 3600 nm in section~\ref{sec:transmission} and the coefficient of thermal expansion (CTE) in section~\ref{sec:cte}. Each of these sections will provide a description of the measurement concept and used setup, followed by the measurement results and a short discussion. The report then ends with a summary and our conclusions in section~\ref{sec:conclusion}.

\section{Index of refraction}\label{sec:ref_index}

Minimum deviation refractometry is the standard method to accurately determine the refractive indices of materials\cite{Tentori1990, Leviton2004_CHARMS}. For this method, a collimated light beam hits the front surface a prism with an apex angle $A$. This prism is then rotated. During the rotation a point of minimum refraction between the incident and outgoing light can be obtained, denoted by $D$. The relation between the refractive index $n$ of the material, the apex angle $A$ and the minimum deviation angle $D$ is given by
\begin{equation}
    n = \frac{\sin{\left( \left( A + D\right) / 2\right)}}{\sin{\left( A / 2\right)}}.\label{eq:mdr_refraction}
\end{equation}
A variation of this method uses the reflection of a light ray on the back surface of the prism, in which case the measurement setup becomes a reflective system and the angle of minimum deviation is the configuration in which the light ray returns exactly in the direction from where it came before encountering the prism. This situation requires that the reflection at the back surface of the prism occurs at a right angle, from which can be deduced that
\begin{equation}
    n = \frac{\sin\theta_i}{\sin A},\label{eq:mdr_reflection}
\end{equation}
where $\theta_i$ is the angle of the ray incident upon the first prism surface for which reflection at the second prism surface occurs.

For practically all classical optical media, the refractive index varies with wavelength. Because the apex angle cannot be a function of wavelength, either $D$ (in transmission) or $\theta_i$ (in reflection) needs to vary with wavelength. In both cases, a measurement setup with a slit, some collimation optics and an accurate rotation stage can be used to precisely measure $D$ or $\theta_i$. When a rotation stage with a high precision angular encoder and a spectrophotometer are used, and a calibration measurement to determine the apex angle $A$ is performed, then the refractive index can be determined through eqs.~\eqref{eq:mdr_refraction} or \eqref{eq:mdr_reflection}.

\subsection{Test setup}
\begin{figure}
    \centering
    \includegraphics[width=0.75\linewidth]{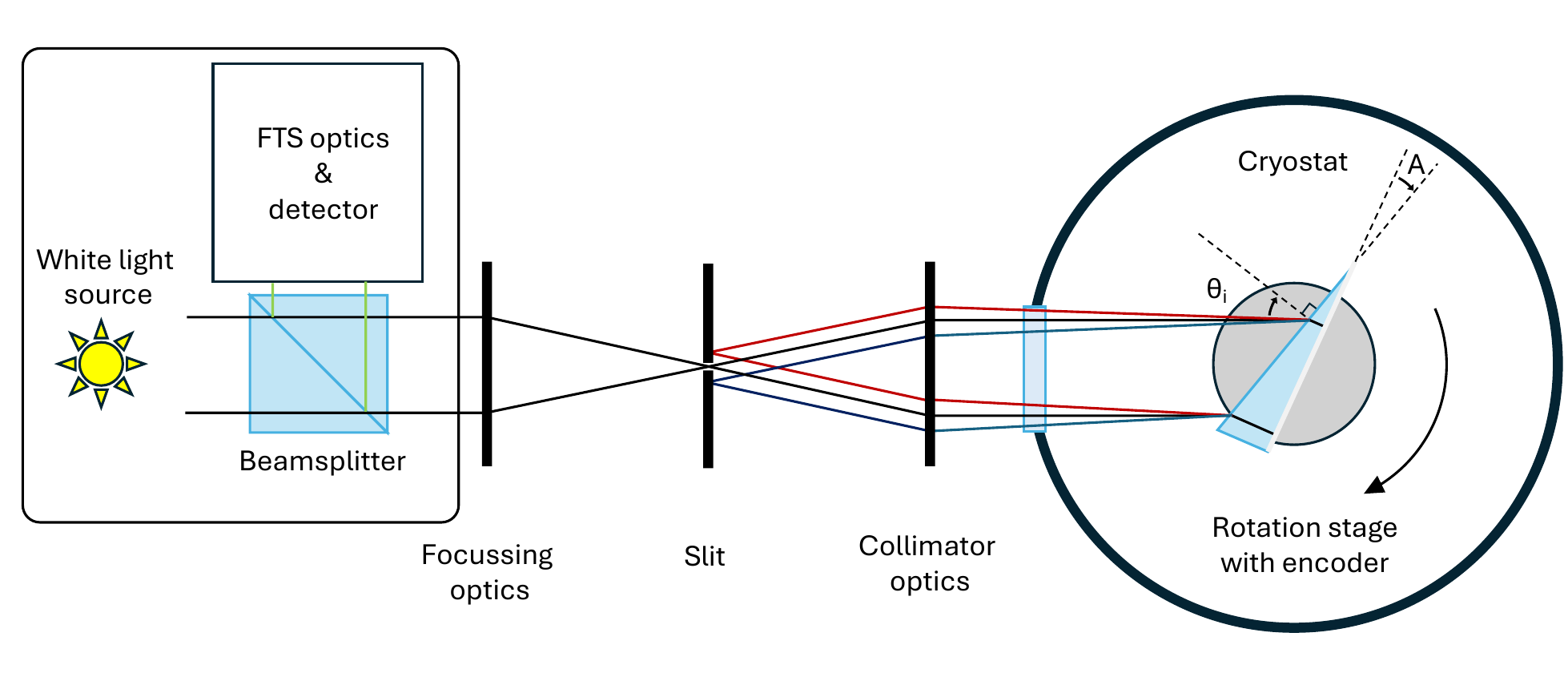}
    \caption{A schematic representation of the used test setup to measure the cryogenic refraction. The FTS contains both the detector and the light source. The light is focused to a focal plane where a slit can be found. The light that passed through the slit is collimated and reflected from the backside of a prism located in the test environment (the cryostat). Then the light returns towards the slit, where only the light with a wavelength such that eq.~\eqref{eq:mdr_reflection} is satisfied can pass through. This then results in a peak in the spectrum measured by the FTS.}
    \label{fig:test_setup_schematic}
\end{figure}

Following the reflective measurement concept described above, we have used a setup designed and built at OMT Solutions. This setup employs a Perkin Elmer Lambda 1050 UV/VIS/NIR spectrophotometer for the spectral measurement and input light. A biconical mirror collimates the light that forms a vertical image of the slit on the prism. The prism itself is mounted between oxygen-free high conductivity copper plates in the cryostat on a rotating stage driven by a stepper motor, accompanied by an ultra-high accuracy angular encoder system. The peak in the spectrum of the light that passes through the slit after retroreflection and which reaches the spectrometer detector is fit to a Gaussian curve, from which the location of the centroid is then determined. A schematic overview of the setup is shown in Fig.~\ref{fig:test_setup_schematic}. 

\subsection{Results}
The refractive index as a function of wavelength can be described by a Sellmeier equation of the following form,
\begin{equation}
    n(\lambda) = \sqrt{1 + \frac{A_1 \lambda^2}{\lambda^2 - B_1^2} + \frac{A_2 \lambda^2}{\lambda^2 - B_2^2} + \frac{A_3 \lambda^2}{\lambda^2 - B_3^2}}.\label{eq:sellmeier}
\end{equation}
In this equation, the wavelength $\lambda$ is given in \textmu m. The temperature dependence of the Sellmeier coefficients $A_i$ are computed as
\begin{equation}
    A_i = C_{0,i} + C_{1,i} \left( \frac{T}{300} \right) + C_{2,i} \left( \frac{T}{300} \right)^2,\label{eq:sellmeier_temperature}
\end{equation}
where the temperature $T$ is provided in Kelvin.

\begin{figure}[!tb]
\centering
\begin{minipage}[t]{.48\textwidth}
\centering
\includegraphics[width=\linewidth]{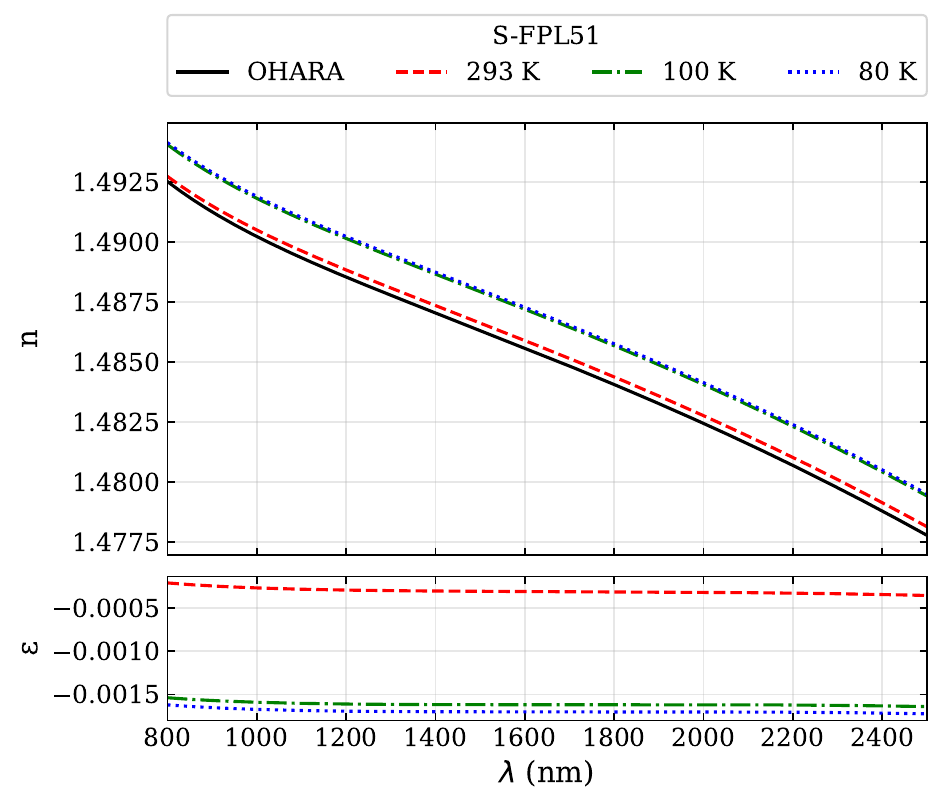}
\end{minipage}%
\hfill
\begin{minipage}[t]{.48\textwidth}
\centering
\includegraphics[width=\linewidth]{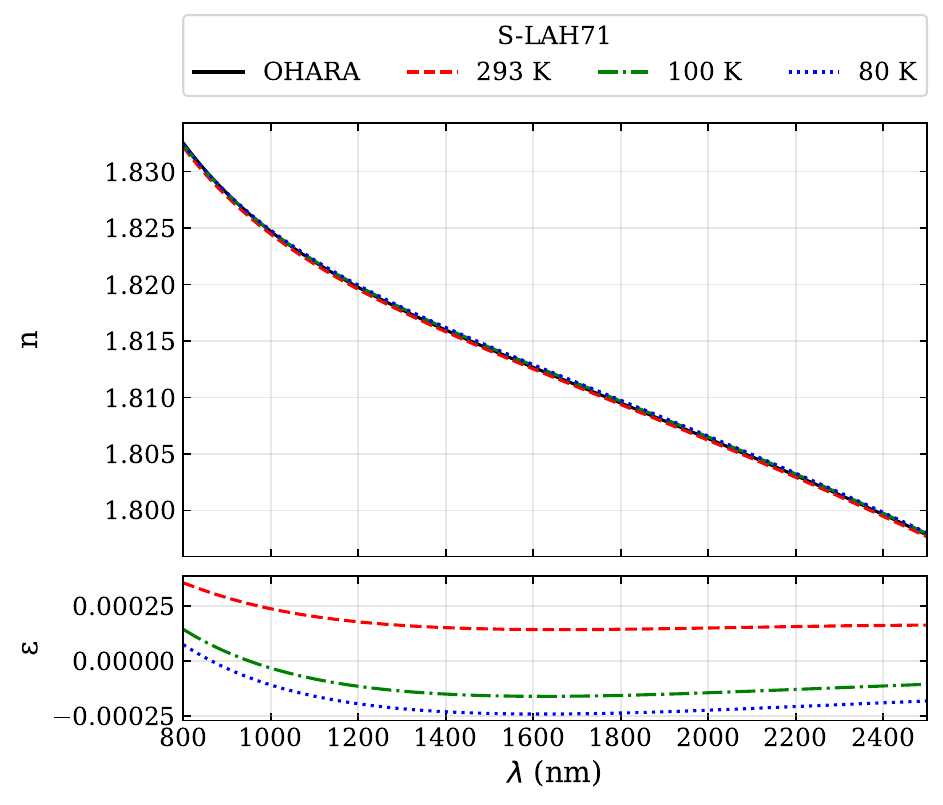}
\end{minipage}%
\caption{The measured refractive index of S-FPL51 (left) and S-LAH71 (right) at 293~K (red dashed line), 100~K (green dash-dot-dashed line) and 80~K (blue dotted line). For reference, also the curve provided in the data sheet is shown in black\cite{OHARA-SFPL51, OHARA-SLAH71}. The bottom panel shows the difference with respect to the data sheet curves. Figure previously presented in Ref.~\citenum{vandenborn2022b}. A tabulated version of the two top panels is provided in Appendix~\ref{app:refractive_index_tables}.}
\label{fig:refractive_index_results}
\end{figure}

The Sellmeier coefficients at 293~K, 100~K and 80~K are presented in Tab.~\ref{tab:sellmeier_coeffs}. The corresponding best fit temperature coefficients are provided in Tab.~\ref{tab:temp_coeffs}. A graphical representation of the measured refractive index of these two materials is given in Fig.~\ref{fig:refractive_index_results}. Between 0.8 and 2.5~\textmu m, the differences in the measured refractive index differ by up to $-1.6\times10^{-3}$ and $3\times10^{-4}$ from the OHARA data for S-FPL51 and S-LAH71. Previously, the cryogenic refractive index of S-FPL51 was measured and presented in Ref.~\citenum{Leviton2013-SFPL51}. Our measurements at 80~K agree to within $1\times10^{-4}$. Another more limited cryogenic measurement had previously been reported in Ref.~\citenum{Yamamuro2006}, to which our measurement agree on the $2\times10^{-4}$ level for the two overlapping data points at 81~K.

\begin{table}[!b]
    \caption{Best fit Sellmeier coefficients of S-FPL51 and S-LAH71 at 293~K, 100~K and 80~K.}
    \centering
    \begin{tabular}{ccSScSSc}
    \toprule
    & Temperature (K) & \multicolumn{1}{c}{$A_1$} & \multicolumn{1}{c}{$A_2$} & \multicolumn{1}{c}{$A_3$} & \multicolumn{1}{c}{$B_1$} & \multicolumn{1}{c}{$B_2$} & \multicolumn{1}{c}{$B_3$}\\
    \midrule
    \multirow{4}{*}{\rotatebox[origin=c]{90}{\centering S-FPL51}} & 293~K & 1.21887 & 1.10497 & - & 0.0819 & 14.211 & - \\[4pt]
    & 100~K & 1.22280 & 1.10944 & - & 0.0819 & 14.211 & - \\[4pt]
    &  80~K & 1.22304 & 1.10935 & - & 0.0819 & 14.211 & - \\[4pt]
    \midrule
    \multirow{4}{*}{\rotatebox[origin=c]{90}{\centering S-LAH71}} & 293~K & 2.2744138 & 0.025379 & 3.26205 & 0.1256 & 0.3739 & 16.73 \\[4pt]
    & 100~K & 2.2779944 & 0.023144 & 3.27669 & 0.1256 & 0.3739 & 16.73 \\[4pt]
    &  80~K & 2.2786028 & 0.022859 & 3.27870 & 0.1256 & 0.3739 & 16.73 \\[4pt]
    \bottomrule
    \end{tabular}
    \label{tab:sellmeier_coeffs}
\end{table}

\begin{table}[!htb]
    \vspace{2\baselineskip}
    \caption{Best fit parameters of the temperature dependent Sellmeier coeficients of S-FPL51 and S-LAH71 for use in eqs.~\eqref{eq:sellmeier} and \eqref{eq:sellmeier_temperature}.}
    \centering
    \begin{tabular}{ccSSc}
    \toprule
    & Coefficient & \multicolumn{1}{c}{j=1} &  \multicolumn{1}{c}{j=2} & \multicolumn{1}{c}{j=3}\\
    \midrule
    \multirow{4}{*}{\rotatebox[origin=c]{90}{\centering S-FPL51}} & $B_j$     & 0.0819    & 14.211    & - \\
    & $C_{0,j}$ & 1.223708  & 1.10792   & - \\
    & $C_{1,j}$ & -0.00157  & 0.0085    & - \\
    & $C_{2,j}$ & -0.00346  & -0.0118   & - \\
    \midrule
    \multirow{4}{*}{\rotatebox[origin=c]{90}{\centering S-LAH71}} & $B_j$     & 0.1256    & 0.3739    & 16.73 \\
    & $C_{0,j}$ & 2.281482  & 0.0216233 & 3.28769 \\
    & $C_{1,j}$ & -0.012134 & 0.004932  & -0.0365 \\
    & $C_{2,j}$ & 0.005014  & -0.001112 & 0.0105 \\
    \bottomrule
    \end{tabular}
    \label{tab:temp_coeffs}
\end{table}



\section{Transmission measurement}\label{sec:transmission}
Both S-FPL51 and S-LAH71 offer very good transmission in the visual and near-infrared wavelengths. In particular, the good transmission of S-LAH71 past 2~\textmu m has been one of the primary reasons for selecting the S-FPL51/S-LAH71 combination as the preferred combination of the IRIS and MICADO atmospheric dispersion correctors\cite{Phillips2016}. 

However, no data on the transmission of either material existed for wavelengths past 2.4~\textmu m. For early trade-off studies and performance estimates this was not a major issue, but for the detailed specification of filter coatings in MICADO, and the known sensitivity of the HAWAII-4RG detectors to radiation outside of the 0.8 to 2.4~\textmu m scientific window, there was a strong desire to learn the spectral behavior of these materials at wavelengths past this window.

\subsection{Test setup}
We have used a Varian FT7000e Fourier Transform Spectrometer (FTS) available at the NOVA optical lab\cite{Navarro2014} to measure the transmission of two witness samples from the same manufacturing run of the S-FPL51 and S-LAH71 material as that will be used for the MICADO ADC. The spectrometer can be equipped with several different light sources and detectors, such that spectra ranging from UV (starting approximately at 250~nm) to mid infrared (ending at approximately 30~\textmu m) can be taken. From earlier experience with the spectrometer we know that the stability of the light source, the detector sensitivity and water vapor contained in atmospheric air can reduce the accuracy of the measurements, but these can be managed with careful planning and calibrations. 

\begin{figure}[!tb]
\centering
\begin{minipage}[t]{.48\textwidth}
\centering
\captionsetup{width=\linewidth}
\includegraphics[width=\linewidth]{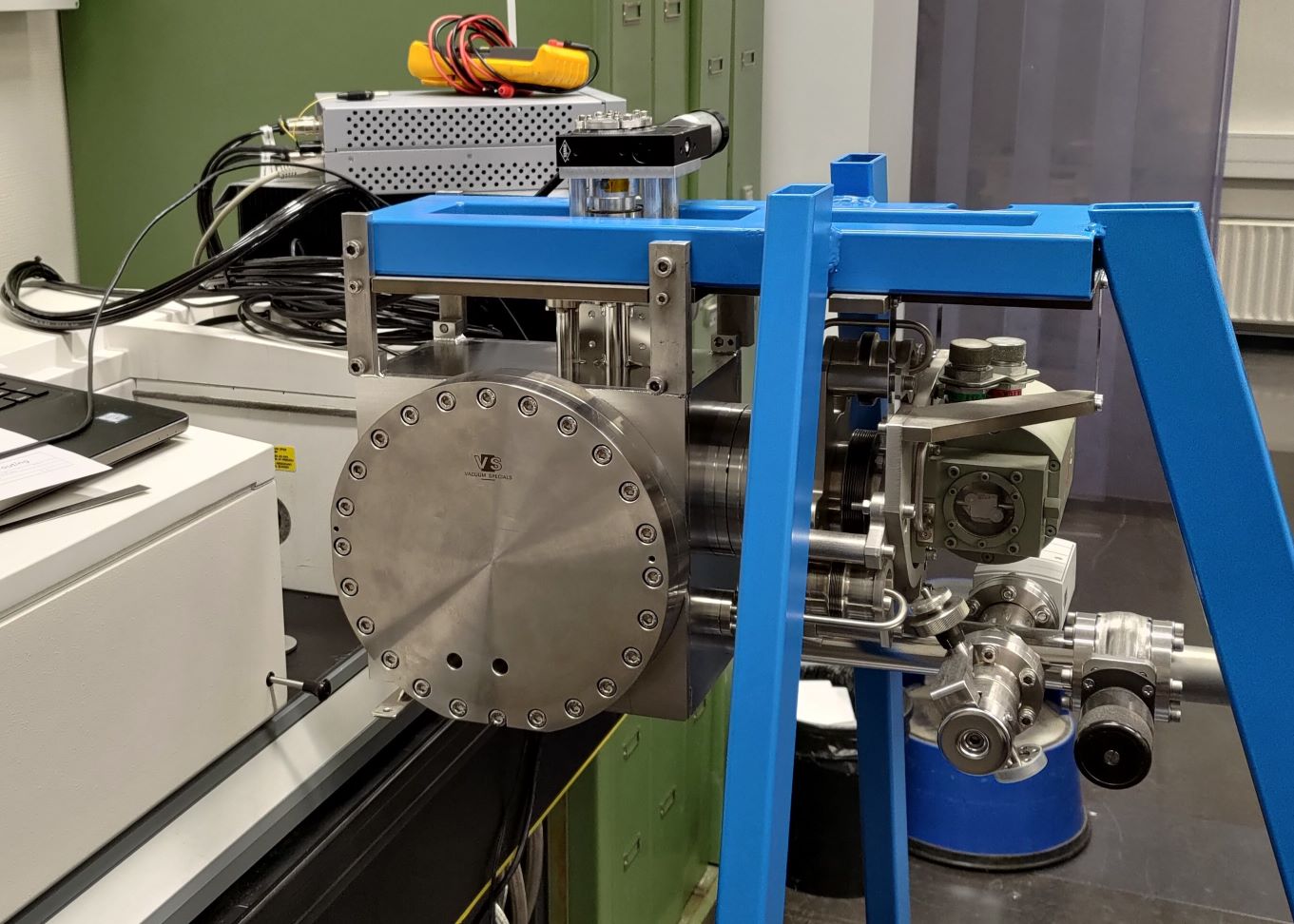}
\captionof{figure}{Picture showing CRiSP before being placed inside the sample compartment of the Varian FT7000e spectrometer at NOVA. Two light tunnels are visible on the the lower half of the cryostat.}
\label{fig:CRiSP_outside}
\end{minipage}%
\hfill
\begin{minipage}[t]{.48\textwidth}
\centering
\captionsetup{width=\linewidth}
\includegraphics[width=\linewidth]{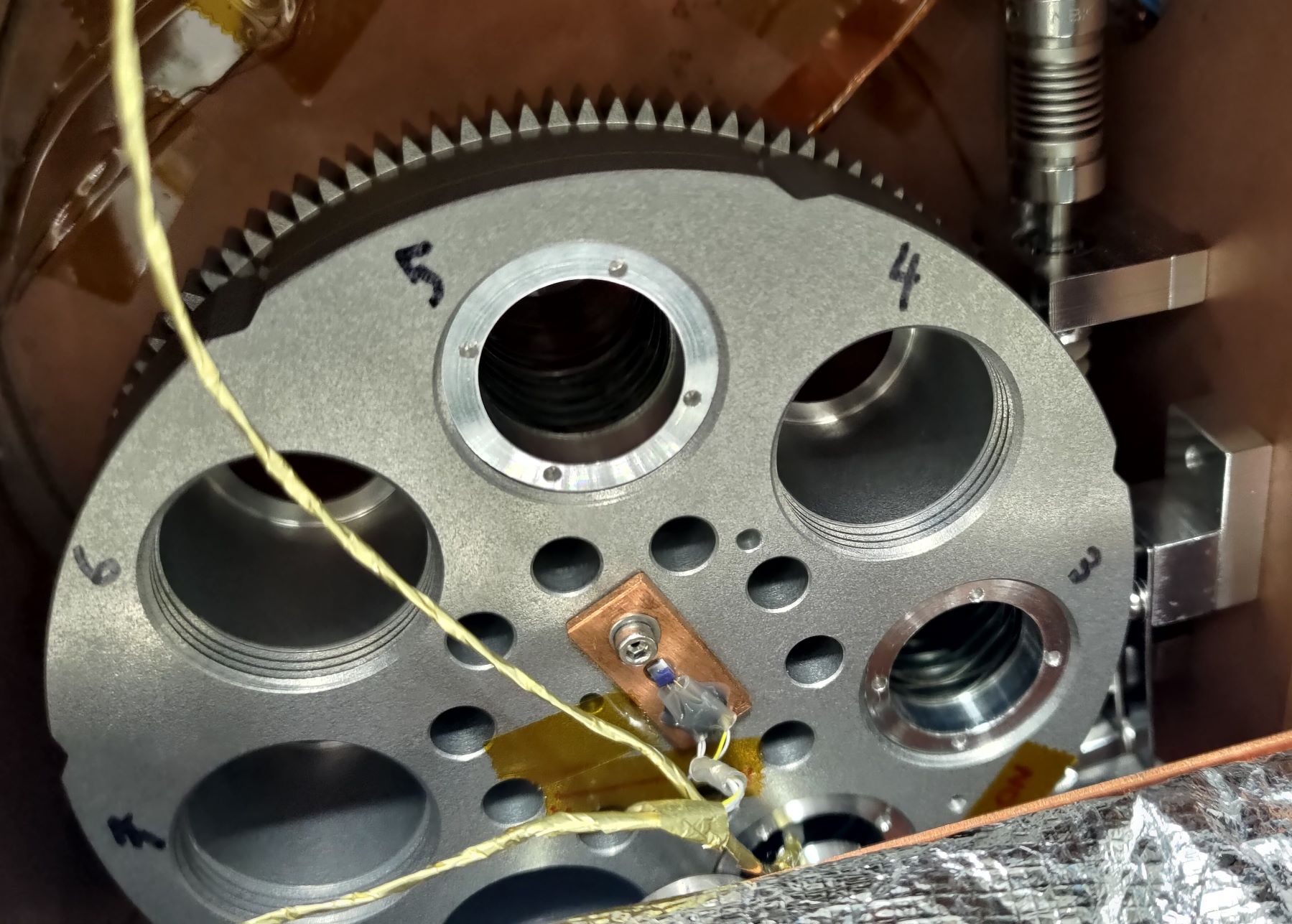}
\captionof{figure}{Picture of the sample wheel within CRiSP. The wheel contains notches to stabilize the position at the sample location. The motor axis connected to the worm gear is visible in the top right.}
\label{fig:CRiSP_inside}
\end{minipage}%
\end{figure}

CRiSP, the CRyostat in SPectrometer, is a custom-built cryostat that was designed and build specifically to fit into the measurement compartment of the spectrometer\cite{Navarro2014}. Located within this cryostat is a sample wheel with slots for seven one-inch samples. The sample wheel can be positioned by a stepper motor on the outside. Because of the worm gear that is used to transfer the rotation of the motor axis to the sample wheel, the gear ratio between the motor axis and the sample wheel is low, requiring the motor to perform 6084° of rotation to move from one sample slot to the next. The sample wheel is pictured in Fig.~\ref{fig:CRiSP_inside} and CRiSP itself is shown in Fig.~\ref{fig:CRiSP_outside}.

CRiSP features two different light tunnels through which a measurement can be performed. One tunnel contains Sapphire windows, which should be used for UV and visible wavelength measurements. It has very good transmission ($>80\%$) between 0.2 and 4 \textmu m, but drops of quickly after 5 \textmu m. For infrared measurements, a different window can be used. This is a Cadmium Telluride (CdTe) window, with a very constant transmission between 60 and 70\% in a spectral range between 1.5 and 28 \textmu m. For this report, we have used the Sapphire window only.

Nominally, the light beam in the measurement compartment of the FTS comes to a focus near the middle of the compartment. Because of the small diameter of the viewing port in CRiSP, the default optical layout loses too much light to be usable. In the past, we have designed a beam compressor that shrinks the incoming beam to a near collimated small beam with a diameter of 9 mm. This allows enough light to pass through CRiSP to be able to perform successful measurements. The beam compressor consists of off-the-shelf components and only the base mounting plate was custom made to fit the FTS interface. 

Unfortunately, the observed shape of the spectrometer light beam was quite different from the expected collimated shape. This is likely due to the light source being an extended source, instead of a point source. Effectively, this means that only a fraction of the total light reaches the detector on the other side of the cryostat resulting in lower than desired signal-to-noise. Nonetheless, it is possible to obtain useful spectral features even at low SNR by stacking multiple scans of the FTS. However, transmission accuracy better than a few percent has been difficult to obtain. For this, a separate measurement of the transmission spectrum is performed with the sample outside of CRiSP, using the standard measurement configuration.

For the transmission measurement of S-FPL51 and S-LAH71, four sample slots were used. One slot in the sample wheel contained the S-FPL51 sample, followed by an open slot for background measurements, then followed by the S-LAH71 sample. A final slot contained glass dummy to which a temperature sensor was glued. This sensor provided the information on the temperature of the real samples. Both glass samples had a diameter of one inch and a thickness of 5~mm and were polished but uncoated.


\subsection{Results}
\begin{figure}
    \centering
    \includegraphics[width=\linewidth]{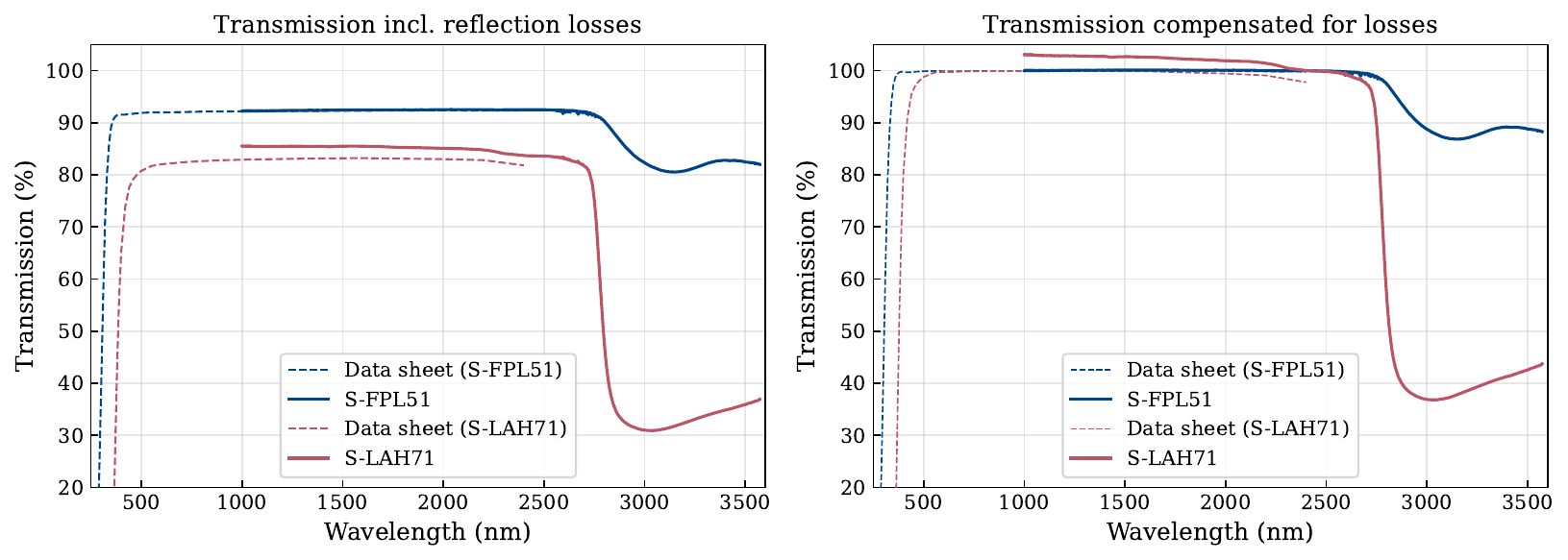}
    \caption{The transmission profile of S-FPL51 is shown in blue. The red line shows the transmission of S-LAH71. The left panel shows the measured transmission, without compensation for any reflection losses of the uncoated samples. With a good anti-reflection coating, a transmission curve more similar to the right panel may be observed. Effectively, the internal transmission is shown in there.}
    \label{fig:transmission_roomtemp}
\end{figure}

Before the cryogenic test was implemented, a regular transmission spectrum was obtained for both samples between 1000 and 3600~nm. This is shown in Fig.~\ref{fig:transmission_roomtemp}. Here, the left panels shows the as measured transmission, including all reflection losses. The right panel uses the refractive index of the materials as measured in section~\ref{sec:ref_index} to compute the Fresnel reflection losses, ignoring any polarization effects and assuming negligible absorption. This information is then used to find the internal transmission of the sample. Due to the instability of the light source in the spectrometer, an internal transmittance above 100 per cent is measured for S-LAH71. Clearly, this cannot be real. The spectral shape of the transmission agrees very well with the manufacturer data sheet, which does stay below 100 per cent transmission. The measurement of S-FPL51 is better behaved and has a maximum internal transmission very close to the expected 100 per cent between 1000 to 2500~nm\cite{OHARA-SFPL51, OHARA-SLAH71}.

Both materials display a decrease in transmission past 2700~nm. This is approximately 12 per cent for S-FPL51 and 60 per cent for S-LAH71. Some weak residual water absorption features can be seen between 2500 to 2700~nm, meaning that the measurement compartment had not completely been flushed by dry nitrogen.

\begin{figure}
    \centering
    \includegraphics[width=\linewidth]{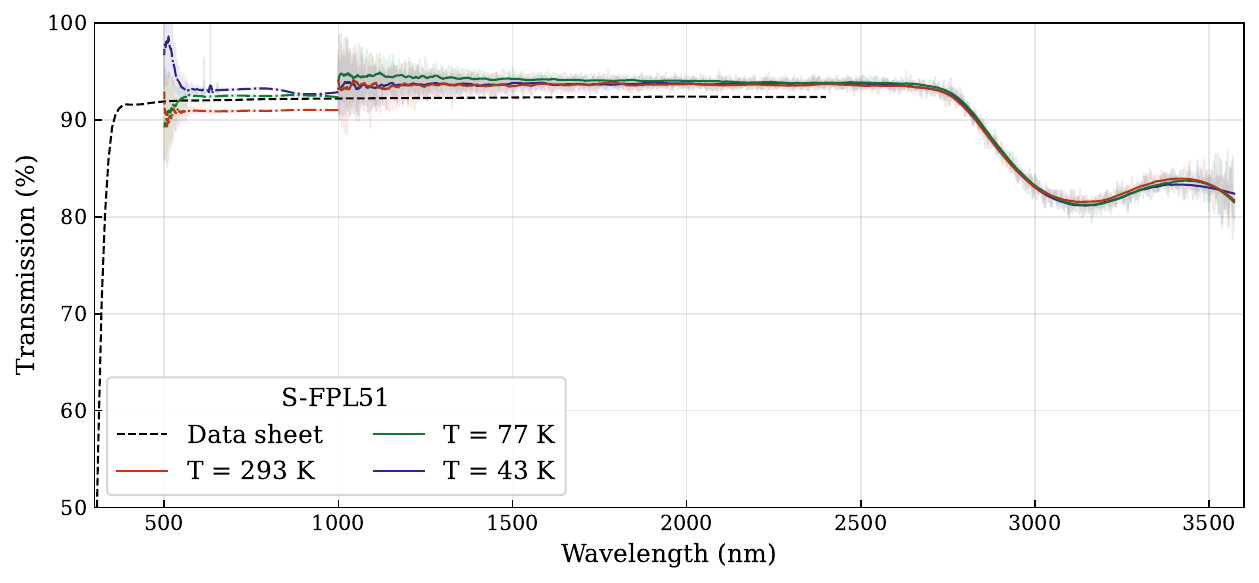}
    \caption{The measured transmission profile of S-FPL51 at 293~K (red), 77~K (green) and 43~K (blue). The black dashed curve provides the data as specified by OHARA. No significant changes in the transmission as a function of temperature can be observed. The amplitude variation is more likely due to the instability of the light source in the spectrometer. The original unsmoothed data is shown at low opacity in the background.}
    \label{fig:transmission_cold_sfpl51}
\end{figure}

\begin{figure}
    \centering
    \includegraphics[width=\linewidth]{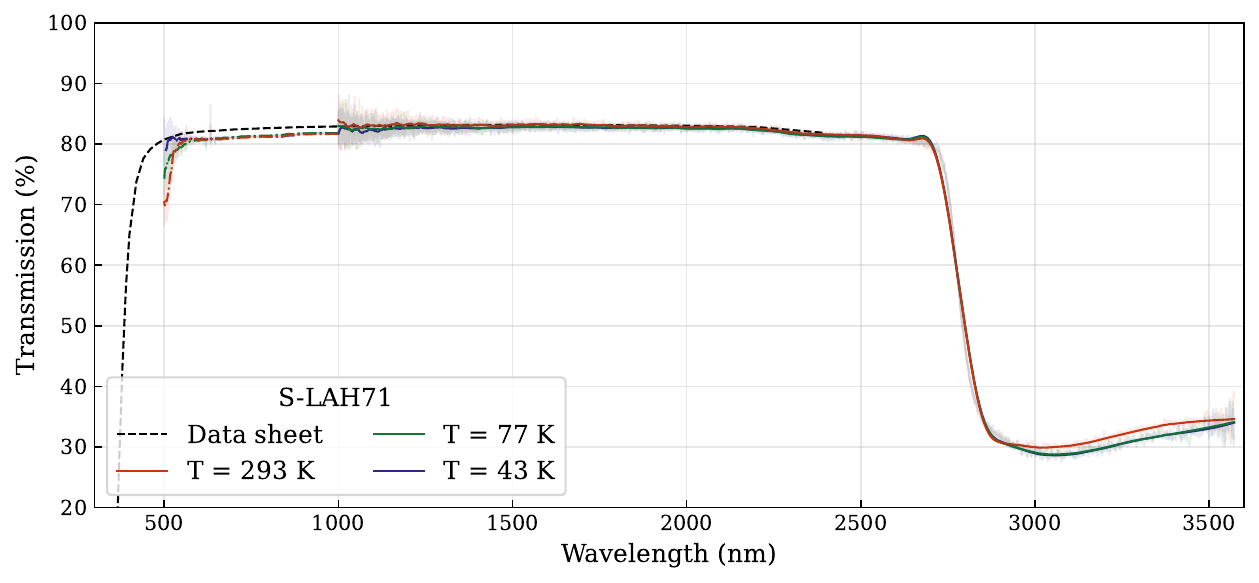}
    \caption{The measured transmission profile of S-LAH71 at 293~K (red), 77~K (green) and 43~K (blue). The black dashed curve provides the data as specified by OHARA. No significant changes in the transmission as a function of temperature can be observed. The amplitude variation is more likely due to the instability of the light source in the spectrometer. The original unsmoothed data is shown at low opacity in the background.}
    \label{fig:transmission_cold_slah71}
\end{figure}

Next, the cryogenic measurements are presented in Figs.~\ref{fig:transmission_cold_sfpl51} and \ref{fig:transmission_cold_slah71} for S-FPL51 and S-LAH71, respectively. Transmission spectra were obtained at room temperature in a vacuum, at 77~K, and at 43~K. Due to a cumbersome alignment of CRiSP in the spectrometer and the limited overall throughput, the signal contained more noise than for the measurement presented in Fig.~\ref{fig:transmission_roomtemp}. Because we did not see any thin absorption features in the measurement at room temperature, nor after a quick inspection of the raw data at cryogenic temperatures, we applied a (non-uniform) Savitzky-Golay filter to reduce the noise\cite{SavitzkyGolay1964}. For this filter, a polynomial of order $p$ is fit to the last $n$ data points at wavenumber $\nu$. A trivial example of this filter is a moving average. Here, we have used $p=3$ and $n=301$.

In the results presented here, some source stability issues occur once again. These give a constant offset of a few percent to the S-FPL51 transmission profile. For all measurements the systematic noise near 500~nm increases significantly, making it unlikely that the variation in the transmission seen in this region is a real signal. Also, the smoothed transmission curve of S-LAH71 shows a minor bump just before the transmission decrease at 2700~nm, which can be attributed to the third order polynomial that the Savitzky-Golay filter is forced to use. The original data does not show this bump. Besides these minor discrepancies, the transmission curve follows the manufactured provided transmission curve well in the overlapping spectral region. We also do not see changes in the transmission profile of any significance at the different temperatures.

\section{Coefficient of Thermal Expansion}\label{sec:cte}

The Coefficient of Thermal Expansion (CTE), denoted by $\alpha$, defines the change in physical size of a sample of size $L_0$ for a decrease in temperature of $\Delta T$. The size at a temperature $T$ is given by

\begin{equation}
    L(T) = L_0 (1 - \alpha \Delta T),
\end{equation}
where the CTE $\alpha$ is defined as
\begin{equation}
    \alpha = \frac{1}{L}\frac{dL}{dT}.\label{eq:cte_definition}
\end{equation}

Perhaps more practical to the engineer is the total change in size of a material relative to room temperature. We can modify the above expression to define an effective CTE (or linearized CTE), where we take the length of the sample at the initial temperature (usually room temperature, 293 K) to be $L_0$ and measure the change in length at temperature $T$. This provides a measure of how much the material will shrink or expand relative to the initial conditions. In other words, 

\begin{equation}
    \alpha_{\mathrm{eff}} = \frac{L(T) - L_0}{L_0} = \frac{\Delta L (T)}{L_0},
\end{equation}
where $T$ is assumed to be in Kelvin. To determine the CTE of a material with good accuracy it is critical to either measure $\Delta L$ very accurately, have a large $L_0$ or both. 

More fundamentally, the CTE of a material behaves non-linear; it changes continuously with temperature. By measuring the CTE value at multiple points it is possible to find an empirical form that describes the CTE value at all temperatures within a range of validity.

We have fit our data to the following three term power series to describe the temperature dependent CTE,
\begin{equation}
    \alpha(T) = A + BT + CT^2.\label{eq:cte_fit}
\end{equation}
And we have fit the linear expansion data to a similar, but slightly different three term power series,
\begin{equation}
    \frac{\Delta L}{L_0} = \left( A + BT + CT^2 \right)\left(T - T_0\right).\label{eq:linear_cte_fit}
\end{equation}

This last equation has a different form than the five term power series used to describe linear expansion by e.g. National Institute of Standards and Technology (NIST)\cite{NIST_Al6061T6,NIST1977_LNG}. Appendix~\ref{app:NISTConversion} provides a conversion between the two descriptions.

\subsection{Test setup}

\begin{figure}[!tb]
\centering
\begin{minipage}[t]{.48\textwidth}
\centering
\captionsetup{width=\linewidth}
\includegraphics[width=\linewidth]{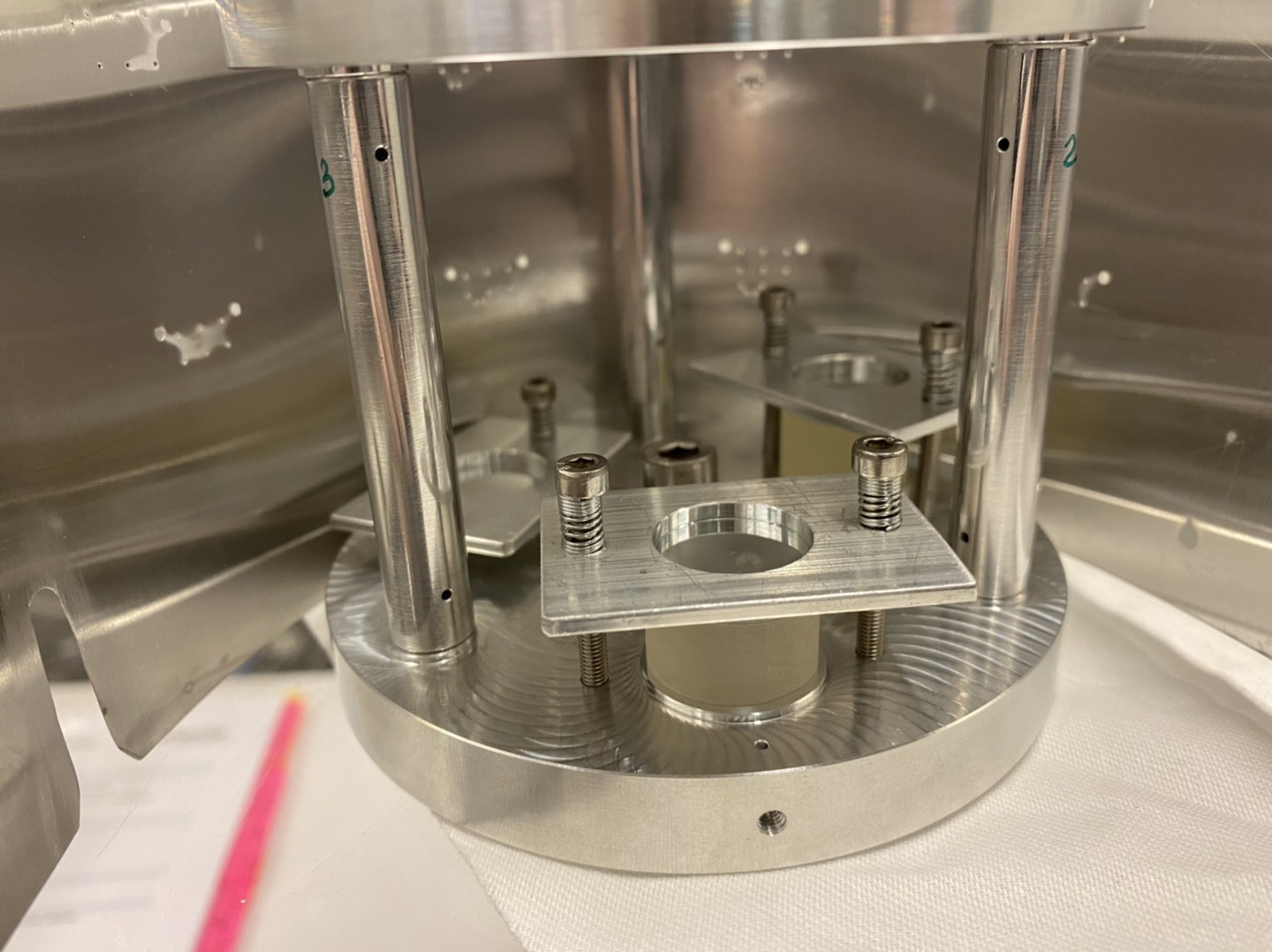}
\captionof{figure}{This picture shows the method in which the samples were placed in the sample holder. The axis on the left only has the reflection mirror and acts as a reference.}
\label{fig:cte_setup_2}
\end{minipage}%
\hfill
\begin{minipage}[t]{.48\textwidth}
\centering
\captionsetup{width=\linewidth}
\includegraphics[width=\linewidth]{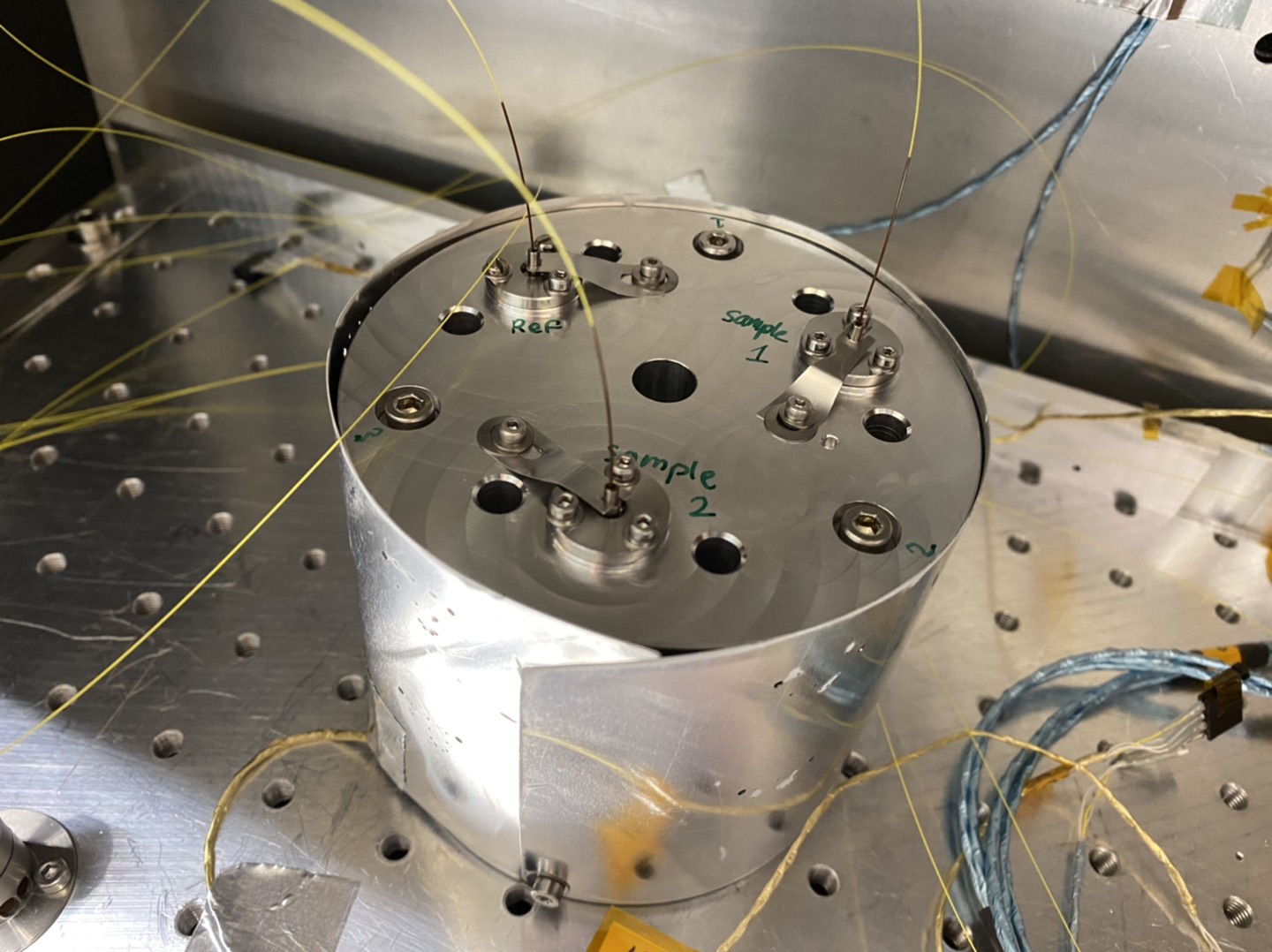}
\captionof{figure}{The sample holder was enclosed in a shield to reduce light leakage and prevent thermal radiation. The three fibers correspond to the three axes of the IDS3010.}
\label{fig:cte_setup_3}
\end{minipage}%
\end{figure}

In our test setup, we have used an interferometric measurement to accurately measure the displacement of the samples as a function of the temperature. The measurement signal was transferred through fibers from the light source to the cryostat, where the light reflected on flat mirrors that had been placed on top of the samples under test. Then, the light returned through the same fibers to the light source unit where the resulting interference signal was processed.

An Attocube IDS3010 interferometric displacement sensor was used for this experiment. The IDS3010 can measure three axes simultaneously at sub-nanometer levels under ideal conditions. A vacuum feedthrough was placed on top of the NOVA Box Cryostat to transfer the signal between the cold vacuum of the cryostat and the warm surroundings of the room. The setup inside the cryostat consisted of a bottom plate on which two samples are placed, see Fig.~\ref{fig:cte_setup_2}. On top of the samples small flat mirrors were placed to reflect the laser light. At least one measurement axis was kept free and acted as the reference signal. This axis only contained the reflective mirror. The reference signal was used to calibrate out any systematic errors, for example from the change in length of the aluminum construction. The top plate was located approximately 60~mm above the bottom plate and housed the sensor heads as illustrated in Fig.~\ref{fig:cte_setup_3}. Alignment was done by tightening or loosening the sensor head housings. After the glass samples were measured, the absolute distance measurement functionality of the IDS3010 was used to find the initial length of the samples at 293~K. Next, a continuous recording of the distance of all three axes was done during cooldown to 40~K and similarly, another recording was done during warmup back to room temperature.

\subsection{Results}
\begin{figure}
    \centering
    \includegraphics[width=0.8\linewidth]{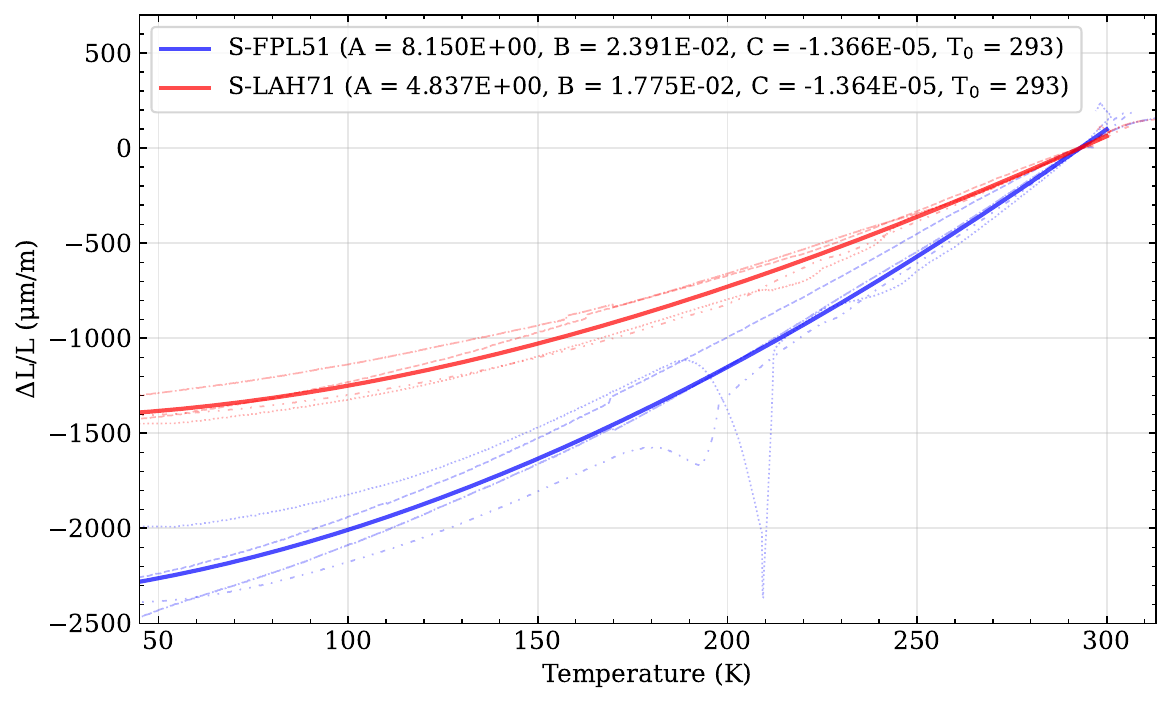}
    \caption{The measured linear expansion of S-FPL51 (blue line) and S-LAH71 (red line). The low opacity lines in the respective colors show the raw data for the two thermal cycles in which the measurement was done. From these lines a temperature lag of approximately $\pm10$~K can be derived. The thick lines are the best fit of the average of the four curves to eq.~\eqref{eq:cte_fit}. The legend provides the coefficients of the linear expansion, which can be converted to follow the NIST convention. This is detailed in appendix~\ref{app:NISTConversion}. This figure in slightly different form was previously presented in Ref.~\citenum{vandenborn2022b}.}
    \label{fig:eff_cte}
\end{figure}

The linear expansion of the two samples is presented in Fig.~\ref{fig:eff_cte}. Here, we show the best fit of eq.~\eqref{eq:linear_cte_fit} to the raw data, consisting of two thermal cycles. A lag existed between the reported temperature and the length of the material, because the temperature sensor was not placed directly on the samples under test, but next to them on the bottom plate. Because we have combined the data from the cooldown and warmup into one curve, this lag should be averaged out and still be a representative fit. Also of note is the large dip in the S-FPL51 data during cooldown between 210 and 190~K. Likely, either the sample, the reference mirror or one of the fiber heads shifted due to a temperature induced stress release. This part of the raw data has been excluded for the best fit, as it is not a real part of the $\Delta L/L$ signal.

From the results, we find a linear shrinkage of 2.1~mm/m and 1.3~mm/m at 77~K relative to room temperature (293~K) for S-FPl51 and S-LAH71, respectively.

We have also tried to derive the temperature dependent CTE values through a numerical evaluation of eq.~\eqref{eq:cte_definition}. Because of small data bins and temperature lag issues, the resulting signal is much noisier. Nonetheless, a clear decreasing trend can be observed with decreasing temperature. Here, we have fit the data to eq.~\eqref{eq:cte_fit}. In addition, we have also numerically applied the curves obtained in Fig.~\ref{fig:eff_cte} as a sanity check and obtained very similar curves. A summary of the best fit coefficients to eqs.~\eqref{eq:cte_fit} and \eqref{eq:linear_cte_fit} is provided in Tab.~\ref{tab:cte_coeffs}. At room temperature the CTE value given for S-FPL51 by OHARA in the data sheets is 13.1~\textmu m/m/K. In good agreement, we find 13.2~\textmu m/m/K from the original data and 14.0~\textmu m/m/K from the application of the $\Delta L/L$ fit. Similarly, for S-LAH71 OHARA reports a value of 7.7~\textmu m/m/K, although we find slightly higher values at 9.9~\textmu m/m/K and 8.9~\textmu m/m/K from the two methods in this case. 

\begin{figure}
    \centering
    \includegraphics[width=0.8\linewidth]{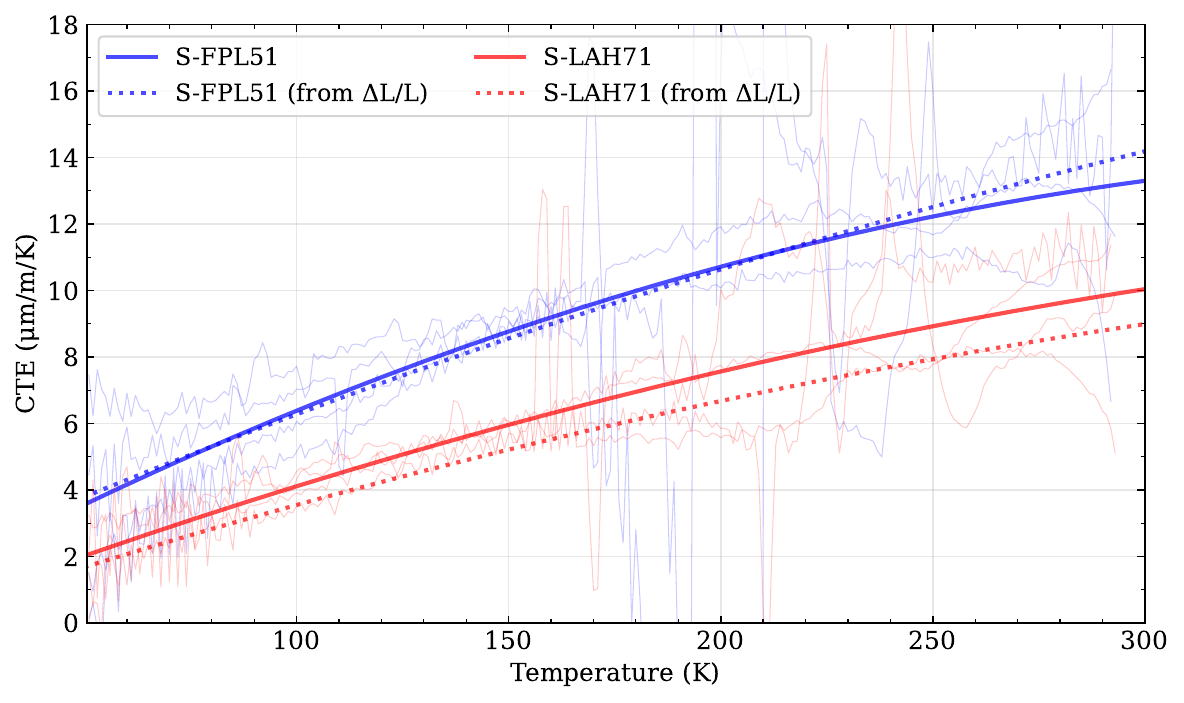}
    \caption{Evaluation of eq.~\eqref{eq:cte_definition} at small temperature bins allows the CTE value to be determined as a function of temperature for both our S-FPL51 sample (blue) and our S-LAH71 sample (red). The low opacity lines in the respective colors show the raw data for the two thermal cycles in which the measurement was done. The thick lines are the best fit of the average of the four curves to eq.~\eqref{eq:cte_fit}. Finally, the dotted lines represent a numerical evaluation of the best fit $\Delta L/L$ curves to obtain the CTE value at any temperature.}
    \label{fig:true_cte}
\end{figure}

\begin{table}[!tb]
    \vspace{\baselineskip}
    \caption{Best fit coefficients for the linear expansion ($\Delta L/L$) and the (temperature dependent) CTE values of S-FPL51 and S-LAH71. The coefficients are used and should be applied to the three term power series given in eqs.~\eqref{eq:cte_fit} and \eqref{eq:linear_cte_fit}. These coefficients provide the CTE value in \textmu m/m/K and the linear expansion in \textmu m/m.}
    \centering
    \begin{tabular}{ccSSSc}
    \toprule
                    &   & \multicolumn{1}{c}{\phantom{abcde}S-FPL51} & \multicolumn{1}{c}{\phantom{abcde}S-LAH71} & Unit \\
    \midrule
    $\alpha(T)$     & A &  3.1082E-01 & -3.1350E-01 & \textmu m m$^{-1}$ K$^{-1}$\\
                    & B &  6.9497E-02 &  4.9173E-02 & \textmu m m$^{-1}$ K$^{-2}$\\
                    & C & -8.7335E-05 & -4.8847E-05 & \textmu m m$^{-1}$ K$^{-3}$\\
    \midrule
    $\Delta L/L$    & A &  8.1504E+00 &  4.8366E+00 & \textmu m m$^{-1}$ K$^{-1}$\\
                    & B &  2.3906E-02 &  1.7752E-02 & \textmu m m$^{-1}$ K$^{-2}$\\
                    & C & -1.3660E-05 & -1.3644E-05 & \textmu m m$^{-1}$ K$^{-3}$\\ 
    \bottomrule 
    \end{tabular}
    \label{tab:cte_coeffs}
\end{table}
\section{Conclusions}\label{sec:conclusion}
In this proceeding, we have reported on our efforts to characterize several properties of OHARA S-FPL51 and S-LAH71 glass at cryogenic conditions. For MICADO, a near-infrared instrument for the upcoming Extremely Large Telescope, these glasses will be used in the atmospheric dispersion corrector and this formed the main motivation for the characterization effort described here. 

Our measurements of the refractive index of S-FPL51 are in good agreement with previous measurements described elsewhere. No earlier cryogenic refractive index measurements of S-LAH71 have been reported and therefore, these results will be useful to future cryogenic instruments that will employ this material. Both S-FPL51 and S-LAH71 show a slight increase in the value of the index of refraction at decreasing temperature. We have provided coefficients to compute the Sellmeier coefficients at any temperature, valid between 80~K and 293~K. 

Next, we have extended the transmission profile of both glasses from the existing transmission data provided by the manufacturer. Both materials show a decrease in transmission past 2700~nm. Especially for S-LAH71 the transmission drop is significant, which will help in relaxing some of the out-of-band blocking requirements of the passband filters in MICADO. When comparing the transmission of the two materials to the transmission at cryogenic temperatures down to 43~K, we have not found any change in transmission or other absorption. 

Finally, we have reported on a measurement of the thermal expansion of the two glasses between 300~K and 45~K. A relatively simple setup based on an interferometric distance measurement was used to determine the linear expansion as well as the change in the coefficient of thermal expansion with temperature. We have found that S-FPL51 shrinks by 1.3 mm/m between room temperature and 77~K, while S-LAH71 shrinks by 2.1 mm/m. The temperature dependent CTE values at room temperature are in rough agreement with the data sheets provided by OHARA, and they showed a decreased value at lower temperatures, although these measurements contained much more measurement noise increasing the uncertainty in the reported values. Nonetheless, the linear expansion values will be most useful in the design of mounting solutions and estimates of stress induced optical aberrations for cryogenic instruments such as ELT-MICADO and TMT-IRIS.


\FloatBarrier
\newpage
\appendix    
\section{Refractive index tables}\label{app:refractive_index_tables}
\vspace{3\baselineskip}
\begin{figure}[!hp]
\centering
\begin{minipage}[t]{.48\textwidth}
\captionsetup{width=\linewidth}
\captionof{table}{The refractive index of S-FPL51 at 80~K, 100~K and 293~K, with an average uncertainty of $2.3\times 10^{-5}$, $2.3\times 10^{-5}$ and $1.7\times 10^{-5}$ at the respective temperatures.}
\centering
    \begin{tabular}{rrrr}
    \multicolumn{4}{c}{\textbf{S-FPL51}}\\
    \toprule
    $\lambda$ (nm) &   T = 80 K &   T = 100 K &   T = 293 K \\
    \midrule
               800 &   1.494145 &    1.494063 &    1.492740 \\
               850 &   1.493493 &    1.493411 &    1.492089 \\
               900 &   1.492913 &    1.492831 &    1.491510 \\
               950 &   1.492388 &    1.492306 &    1.490987 \\
              1000 &   1.491905 &    1.491823 &    1.490505 \\
              1050 &   1.491456 &    1.491374 &    1.490057 \\
              1100 &   1.491031 &    1.490949 &    1.489634 \\
              1150 &   1.490626 &    1.490544 &    1.489229 \\
              1200 &   1.490235 &    1.490153 &    1.488839 \\
              1250 &   1.489854 &    1.489772 &    1.488460 \\
              1300 &   1.489482 &    1.489399 &    1.488088 \\
              1350 &   1.489113 &    1.489031 &    1.487721 \\
              1400 &   1.488748 &    1.488666 &    1.487357 \\
              1450 &   1.488384 &    1.488302 &    1.486994 \\
              1500 &   1.488019 &    1.487937 &    1.486630 \\
              1550 &   1.487653 &    1.487571 &    1.486265 \\
              1600 &   1.487284 &    1.487202 &    1.485897 \\
              1650 &   1.486911 &    1.486829 &    1.485525 \\
              1700 &   1.486534 &    1.486451 &    1.485149 \\
              1750 &   1.486151 &    1.486069 &    1.484768 \\
              1800 &   1.485763 &    1.485681 &    1.484381 \\
              1850 &   1.485369 &    1.485286 &    1.483988 \\
              1900 &   1.484968 &    1.484885 &    1.483588 \\
              1950 &   1.484559 &    1.484477 &    1.483180 \\
              2000 &   1.484143 &    1.484061 &    1.482766 \\
              2050 &   1.483719 &    1.483637 &    1.482343 \\
              2100 &   1.483287 &    1.483204 &    1.481912 \\
              2150 &   1.482847 &    1.482764 &    1.481473 \\
              2200 &   1.482397 &    1.482314 &    1.481025 \\
              2250 &   1.481939 &    1.481856 &    1.480567 \\
              2300 &   1.481471 &    1.481388 &    1.480101 \\
              2350 &   1.480994 &    1.480911 &    1.479625 \\
              2400 &   1.480507 &    1.480424 &    1.479140 \\
              2450 &   1.480010 &    1.479927 &    1.478645 \\
              2500 &   1.479503 &    1.479420 &    1.478139 \\
    \bottomrule
    \end{tabular}
    \label{tab:refractive_index_SFPL51}
\end{minipage}%
\hfill
\begin{minipage}[t]{.48\textwidth}
\captionsetup{width=\linewidth}
\captionof{table}{The refractive index of S-LAH71 at 80~K, 100~K and 293~K, with an average uncertainty of $2.4\times 10^{-5}$, $1.5\times 10^{-5}$ and $1.8\times 10^{-5}$ at the respective temperatures.}
\centering
    \begin{tabular}{rrrr}
    \multicolumn{4}{c}{\textbf{S-LAH71}}\\
    \toprule
    $\lambda$ (nm) &   T = 80 K &   T = 100 K &   T = 293 K \\
    \midrule
               800 &   1.832463 &    1.832394 &    1.832182 \\
               850 &   1.830116 &    1.830044 &    1.829811 \\
               900 &   1.828107 &    1.828032 &    1.827784 \\
               950 &   1.826358 &    1.826282 &    1.826022 \\
              1000 &   1.824813 &    1.824736 &    1.824466 \\
              1050 &   1.823429 &    1.823351 &    1.823073 \\
              1100 &   1.822172 &    1.822094 &    1.821809 \\
              1150 &   1.821018 &    1.820939 &    1.820650 \\
              1200 &   1.819946 &    1.819866 &    1.819573 \\
              1250 &   1.818940 &    1.818860 &    1.818563 \\
              1300 &   1.817987 &    1.817907 &    1.817607 \\
              1350 &   1.817077 &    1.816996 &    1.816695 \\
              1400 &   1.816201 &    1.816120 &    1.815817 \\
              1450 &   1.815351 &    1.815271 &    1.814967 \\
              1500 &   1.814523 &    1.814443 &    1.814138 \\
              1550 &   1.813711 &    1.813630 &    1.813326 \\
              1600 &   1.812911 &    1.812830 &    1.812526 \\
              1650 &   1.812119 &    1.812038 &    1.811734 \\
              1700 &   1.811332 &    1.811251 &    1.810948 \\
              1750 &   1.810548 &    1.810467 &    1.810165 \\
              1800 &   1.809763 &    1.809683 &    1.809382 \\
              1850 &   1.808977 &    1.808897 &    1.808597 \\
              1900 &   1.808188 &    1.808108 &    1.807810 \\
              1950 &   1.807393 &    1.807314 &    1.807017 \\
              2000 &   1.806592 &    1.806513 &    1.806218 \\
              2050 &   1.805784 &    1.805705 &    1.805412 \\
              2100 &   1.804967 &    1.804888 &    1.804597 \\
              2150 &   1.804140 &    1.804061 &    1.803772 \\
              2200 &   1.803302 &    1.803224 &    1.802938 \\
              2250 &   1.802454 &    1.802376 &    1.802092 \\
              2300 &   1.801593 &    1.801515 &    1.801234 \\
              2350 &   1.800719 &    1.800642 &    1.800364 \\
              2400 &   1.799833 &    1.799756 &    1.799480 \\
              2450 &   1.798932 &    1.798856 &    1.798583 \\
              2500 &   1.798017 &    1.797941 &    1.797672 \\
    \bottomrule
    \end{tabular}
    \label{tab:refractive_index_SLAH71}
\end{minipage}%
\end{figure}
\FloatBarrier
\newpage
\section{Conversion to the NIST equation}\label{app:NISTConversion}
The standard equation describing linear expansion as used by the NIST, e.g. for Aluminium 6061-T6\cite{NIST_Al6061T6}, is a five term power series,

\begin{equation}
    y = a + bT + cT^2 + dT^3 + eT^4, \label{eq:NIST}
\end{equation}

where y is the linear expansion $\Delta L/L_0$ and $T$ is given in Kelvin. We can convert the equations and coefficients provided in the previous sections into this form.

From the definition of thermal expansion, we can rewrite
\begin{equation}
    L(T) = L_0 \left( 1 - \alpha \Delta T\right)
\end{equation}
into

\begin{equation}
    \alpha \Delta T = \frac{L_0 - L(T)}{L_0} = (A + BT + CT^2)(T-T_0).
\end{equation}

Writing this out leads to the form of the NIST equation, where
\begin{equation}
    \begin{aligned}
        a &= -AT_0,\\
        b &= A - BT_0,\\
        c &= B - CT_0,\\
        d &= C,\\
        e &= 0.
    \end{aligned}
\end{equation}
The converted coefficients for S-FPL51 and S-LAH71 are provided in Tab.~\ref{tab:NISTCoefficients} below. As in the main text, we have presented the thermal expansion in units of \textmu m/m, but a conversion to the unitless m/m, which is used regularly too, is only a factor 10\textsuperscript{-6} away.
\vspace{3\baselineskip}

\begin{table}[!h]
    \caption{Coefficients of the linear expansion of S-FPL51 and S-LAH71, converted into the five term power series format used by e.g. NIST. This makes it easier to directly compare the measured expansion to other materials.}
    \centering
    \begin{tabular}{lcccccc}
    \toprule
    Material & \multicolumn{1}{c}{$T_0$ (K)} &  \multicolumn{1}{c}{a} & \multicolumn{1}{c}{b} & \multicolumn{1}{c}{c} & \multicolumn{1}{c}{d} & \multicolumn{1}{c}{e} \\
    \midrule
     S-FPL51 & 293 & -2.3881E+03 & 1.1460E+00 & 2.7908E-02 & -1.3660E-05 & 0 \\
     S-LAH71 & 293 & -1.4171E+03 & -3.6473E-01 & 2.1750E-02 & -1.3644E-05 & 0 \\
    \bottomrule
    \end{tabular}
    \label{tab:NISTCoefficients}
\end{table}

\newpage
\bibliography{report} 
\bibliographystyle{spiebib} 

\end{document}